\documentclass[final,5p,times,twocolumn]{elsarticle}

\usepackage{amssymb}
\usepackage{amsmath}
\usepackage{hyperref}
\usepackage{xcolor}

\usepackage[latin]{armtex}

\hypersetup{
    colorlinks,
    linkcolor={red!50!black},
    citecolor={blue!50!black},
    urlcolor={blue!80!black}
}

\journal{Physica C}

\begin{document}

\begin{frontmatter}



\title{Something about the Mechanism of Induction Phenomena: \\ The forgotten work of Berta de Haas-Lorentz on diamagnetism in superconductors}


\author[dqmp]{Giulia Venditti} 
\author[leiden]{Carlo Beenakker} 
\author[dqmp,leiden]{Louk Rademaker} 

\affiliation[dqmp]{organization={Department of Quantum Matter Physics, University of Geneva},
            addressline={24 Quai Ernest-Ansermet}, 
            city={Geneva},
            postcode={1211}, 
            country={Switzerland}}
\affiliation[leiden]{organization={Institute-Lorentz for Theoretical Physics, Leiden University},
            addressline={P.O. Box 9506}, 
            city={Leiden},
            postcode={2300 RA}, 
            country={The Netherlands}}

\begin{abstract}
In 1925, Dr. Geertruida Luberta ``Berta'' de Haas-Lorentz published the paper ``{\em Iets over het mechanisme van inductieverschijnselen}" in the journal {\em Physica}. Her paper was the first to discuss perfect diamagnetism of superconductors, eight years before the discovery of the Meissner effect, when the essential difference between the two phenomena was not understood. Unfortunately, her work was almost forgotten by the scientific community. To counter this, we translate her seminal 1925 paper from Dutch into English. We provide an overview of the life of Dr. De Haas-Lorentz, and comment on her pioneering contribution to the theory of superconductivity.
\end{abstract}







\end{frontmatter}




\tableofcontents


\section{Introduction}

Hundred years ago female physicists were rare, and it was even rarer that their contributions were noticed or honored. Nevertheless, many women did contribute to the progress of physics, and one of them was Berta~de~Haas-Lorentz. While some of her work is known, for example on noise and Brownian motion in conductors, her contribution to the field of superconductivity has been mostly ignored. A major reason that this work was not recognized, is that it was published in Dutch-language journal. 

On the centenary of the publication of ``Something about the mechanism of induction phenomena''\cite{Berta1925-SC}, we translated her work and put her results in the historical and scientific context.

This work is organized as follows:
In Section~\ref{sec:bio} we describe her biography, and her major scientific contributions.
In Section \ref{sec:paper} we present the translation, from Dutch to English, of her paper \cite{Berta1925-SC}.
Finally, in Section \ref{sec:context} we provide a scientific analysis of her paper in its historical perspective.


\section{The life of Berta de Haas-Lorentz}
\label{sec:bio}

\subsection{Life}
Geertruida Luberta ``Berta" de Haas-Lorentz was a Dutch theoretical physicist, born in Leiden on 22 November 1885.
As the reader might have noticed, her surname is composed of two well-established names in the field. 
Her father was Hendrik Lorentz, who won the 1902 Nobel Prize together with Pieter Zeeman, and is known for many contributions to physics including the Lorentz force and the Lorentz transformation. Hendrik had a good relationship with his kids; he taught Berta math and how to read maps, and learned all his kids Latin names of flowers when on hikes. Berta's mother, Aletta Kaiser, was a women’s rights activist, involved with the local women's suffragist movement. 

In 1910, Berta married Wander Johannes de Haas, an experimental physicist who later became famous for the Einstein-de Haas effect and his discovery of quantum oscillations in large magnetic fields (the De Haas–Van Alphen and Shubnikov–de Haas effects).
Wander pursued his PhD research in the group of Kamerlingh Onnes, who was the first to liquefy helium in 1908 and discovered superconductivity in 1911. Berta worked as an assistant in Kamerlingh Onnes' lab from 1908 to 1910, during which time she helped Wander with his experiments on pressurized hydrogen.\footnote{Wander de Haas thanks his wife in the acknowledgments of his PhD thesis: {\em ``With pleasure I convey my thanks towards Mr. G. Holst and Madame G. L. de Haas-Lorentz for their carefulness in determining hydrogen temperatures."}}

Following her work in the lab, Berta switched to theory and joined her father's group to do her PhD.  At the time it was quite unusual to have female PhD students, let alone in physics. Nevertheless, Hendrik Lorentz  supervised in total 4 female PhD students! In fact, they were also the first four female PhD students in physics in Leiden~\cite{vandelft}. 
After them, the University of Leiden would have to wait until 1938 for a fifth woman to earn a physics PhD (Anna Keesom, daughter of Willem Keesom).
During her studies, Berta was also an active member of the {\em VVSL} ({\em Vereniging van Vrouwelijke 
Studenten te Leiden}), the first female student society in Leiden.\footnote{On the history of this society, see \url{https://www.universiteitleiden.nl/en/dossiers/herstory/associations/vvsl}.}

Before completing their PhDs, the De Haas couple moved to Berlin in September 1911. There, Wander
started working for H.~E.~J.~G. du Bois, while Berta gave birth to their first child, son Albert. In the mean-time she continued working on research -- for example, while she was pregnant, her father suggested in a letter: “Now that you have some extra time to work, shall you focus on the Brownian motion?”\cite{vandelft}

On July 11$^\text{th}$, 1912, Wander de Haas defended his PhD thesis on pressurized hydrogen~\cite{wanderthesis}. 
Berta de Haas-Lorentz earned her doctorate few months later on September 24$^\text{th}$, with a dissertation entitled {\em Over de theorie van de Brown'sche beweging en daarmede verwante verschijnselen} (On the theory of Brownian motion and related phenomena)~\cite{Berta1912-thesis}.

When World War I broke out in 1914, Wander and Berta and their first two kids were still in Berlin. The Netherlands remained neutral throughout the war, and in 1915 Hendrik Lorentz came personally to Berlin to bring his daughter, son-in-law and grandchildren back to The Netherlands. After several positions throughout the country, Wander de Haas finally came back to Leiden in 1924, when he inherited the Kamerlingh Onnes laboratory, pursuing the challenge of studying the phenomenon of superconductivity and properties of materials in a high magnetic field.

Meanwhile, Berta continued to publish theoretical research. Following her fathers strict neutrality policy after the war, she mostly published in Dutch, and not in the more common -- but politically charged -- languages German or French. 

In 1957 she published a book about the life of her father, with personal anecdotes and contributions from famous physicists like Einstein, Ehrenfest and Casimir. Wander died in 1960, and Berta died in 1973 in Leiden, leaving behind four children.

\subsection{Research}

After Albert Einstein's paper on Brownian motion came out in 1905~\cite{Einstein1905molekularkinetischen},
 Berta was one of the first to work on possible applications of the theory to other domains.
 Her PhD topic was mentioned in the lecture by Solvay himself at the first Solvay conference in 1911.
She carried out analysis of electron fluctuations as Brownian particles, in times when the technology to observe electron fluctuations was still about to be developed and the quantum revolution was hardly started: the discovery of the electron dates back to 1897 (only 15 years before her dissertation) while the Bohr atomic model will be formulated only one year after, in 1913.
Among her results, she anticipated the Johnson–Nyquist noise~\cite{Dorfel2012early}, 
 which was studied only in the next decade and experimentally observed and understood 12 years after her dissertation was published~\cite{Nyquist128, Johnson1928}.

She also had a role in the understanding of magnetism. 
Together with her husband, she worked on what later would be known as the "Einstein-de Haas effect"~\cite{einstein-dehaas1915experimental}.
The Einstein-de Haas effect was indeed the experimental observation of Ampère's {\em molecular currents} 
, or, in more contemporary words, the observation that, as a consequence of momentum conservation, a magnetic moment in a ferromagnet can induce a mechanical moment (a torque) in the material. In Ref.~\cite{DeHaasCouple1917experiment},
the couple showed that an earlier claim by Maxwell that no torque was observable, was related to the small numbers involved in Maxwell's setup caused by the fact that the system was not at resonance frequency. This proved, once more, the correctness of Einstein-de Haas effect.
It is worth noting that the Einstein-de Haas effect and the de Haas couple paper were both written during World War I when Wander and Berta were still in Berlin. 

One of her most remarkable contributions, however, was in the newborn field of superconductivity.

\section{Something about the mechanism of induction phenomena}
\label{sec:paper}

{\em This section contains a translation of the original 1925 paper "Iets over het mechanisme van inductieverschijnselen". \cite{Berta1925-SC}.

Note that the scientific language has changed significantly since 1925, also due to our better understanding of quantum mechanics and electromagnetism.
In particular, we need to remark on two phrases often used in the paper.

``Moleculaire kringstromen" is translated as ``molecular circular currents".
This phrase has been used in the study of the origins of magnetism, also by Wander de Haas in his publications of the Einstein-De Haas effect, to describe currents causing a magnetic moment. The word ``molecular" does not refer to molecules in the modern chemical sense, it is just meant as ``very small". A modern version of the phrase could be ``bound currents", but we chose to keep the original language of Berta.

On the other hand, we translate her phrase ``krachtlijnen" (literally: force lines) as ``magnetic field lines", consistent with the modern nomenclature.}

\

\

\begin{center}
{\bf Iets over het mechanisme van inductieverschijnselen}\\
(Something about the mechanism of induction phenomena) \\
{\em by Geertruida Luberta de Haas-Lorentz}
\end{center}

{\bf 1.} Among the investigations into superconductors, which are carried out in the Leiden laboratory of prof. de Haas,
certainly those into a change of the magnetic
properties of superconducting materials would included, 
if such an investigation were not made extremely difficult by the superconducting condition itself.
After all, it will never be possible
to bring magnetic fields lines through a superconductor without the generated induction currents creating an even large oppositely directed
magnetic field, which remains, since the induction currents are not extinguished by any resistance.

While considering this, he posed the question: contrary to what we stated above, how can we imagine that molecular circular currents could induce magnetic field lines, even though these currents experience no resistance?

\vspace{10pt} 

\noindent {\bf 2.} Prof. Ehrenfest pointed out the essential difference between resistanceless molecular circular currents, and the currents generated through induction in a superconductor. Without a doubt, this has to be the origin of their different behavior with regard to magnetic field lines.

Consider the kinetic energy of an electric current carried by electrons. This consists of three\footnote{The original text writes `drie' (three), but only two parts are mentioned afterwards.} 
parts:
\begin{enumerate}
    \item[A.] The energy $T_L = \frac{1}{2} L i^2$, located in the magnetic field associated with the current,
    \item[B.] The energy, located in the immediate vicinity of the electrons, which is commonly expressed as the mechanical kinetic energy $T_K = \sum \frac{1}{2} m v^2$ where $m$ is the mass of the electron.
\end{enumerate}

The effect of both these contributions to the kinetic energy is best clarified with a simple example.

We imagine a ring current (with current strength $i$) and at a adjustable distance a coil (with current strength $I$). Since both $T_K$ and $T_L$ are proportional to $i^2$, we can write for the energy of the ring
\begin{equation}
    \frac{1}{2} (L^1 + L)i^2
\end{equation}
and equally for the coil
\begin{equation}
    \frac{1}{2} L I^2.
\end{equation}
The energy of the coil and the ring together is
\begin{equation}
    \frac{1}{2} (L^1 + L)i^2 + M i I + \frac{1}{2} L I^2.
\end{equation}
When there acts no electromotive force in the ring, we have
\begin{equation}
    \frac{d}{dt} \left\{ (L^1 + L)i + MI \right\} = 0
\label{eq:perfectdiamagnet}
\end{equation}
or equivalently
\begin{equation}
     (L^1 + L)i + MI = {\rm constant; \; }
     {\rm for\,  example}\footnote{In Dutch she writes here `b.v.', which is the common acrynom for `bijvoorbeeld', meaning {\em for example}.} = 0
\label{eq:meissner?}
\end{equation}
such that
\begin{equation}
    i = - \frac{M}{L^1 + L} I.
\end{equation}
The number of field lines that goes through the ring is
\begin{eqnarray}
    N & = & Li + MI \\
    & = & MI \left\{ 1 - \frac{L}{L^1+L} \right\}.
\end{eqnarray}
We can now very clearly distinguish two limiting cases:
\begin{enumerate}
    \item[I.] The case
    \begin{equation}
    \begin{split}
         \frac{L^1}{L} \ll 1, &\; {\rm or } \; \frac{T_L}{T_K} \gg 1,
    \\
    &N \cong 0
    \label{eq:caseI-SC}
    \end{split}
    \end{equation}
    which means that of all the magnetic field lines the coil generates, practically none of them pass through ring. This one extreme case is relevant for the superconducting ring we mentioned above;
        
    \item[II.] The case
    \begin{equation}
    \begin{split}
         \frac{L^1}{L} \gg 1, &\; {\rm or } \; \frac{T_L}{T_K} \ll 1,
    \\
    &N \cong MI
    \end{split}
        \label{eq:caseII-m}
    \end{equation}
    which means that all the field lines originating from the coil go through the ring. This other extreme case is likely occuring in molecular circular currents.
\end{enumerate}

\vspace{10pt}

\noindent {\bf 3.} I thought it would be interesting to verify the ratio $T_L/T_K$ for some given situation, and also to investigate whether we can realize an intermediate situation where the influence of both $T_L$ and $T_K$ can be felt. For such an intermediate regime, $T_L/T_K$ should not differ too much from 1.

We consider first the simplest imaginable situation: $n$ electrons who at equal distance move with a velocity $v$ along the circumference of a circle of radius $a$. The field energy is (we will use electrostatic units throughout):
\begin{equation}
    T_L = \frac{n e^2 v^2}{3 a c^2}
        \sum_{h=1}^{n-1} 
        \frac{1 - 2 \sin^2 \frac{h\pi}{n}}{\sin \frac{h\pi}{n}}
\end{equation}
Replacing the sum by an integral (with variable $h\pi/n$, with lower bound $\pi/n$), we find
\begin{equation}
    T_L = \frac{n^2 e^2 v^2}{2\pi a c^2}
    \left[ - \log \left( \tan \frac{\pi}{2n} \right) - 2 \right]
\end{equation}
which becomes in the limit of very large $n$
\begin{equation}
    T_L = \frac{n^2 e^2 v^2}{2\pi a c^2} \log n.
\end{equation}
The kinetic energy is
\begin{equation}
    T_K = \frac{1}{2} n m v^2
    = \frac{n e^2 v^2}{12 \pi R c^2}
\end{equation}
where $R$ is the radius of the electron, and assuming the charge of the electron is distributed over its surface.

The ratio is
\begin{equation}
    \frac{T_L}{T_K} = 
    \frac{6 R \, n \log n}{a}
    = 12 \pi \frac{R}{D} \log n
    = 12 \pi \frac{R}{D} \log \frac{2\pi a}{D}
\end{equation}
where $D$ is the average distance between two electrons.

Since $R = 10^{-13}$
and $D$ can not be smaller than something on the order of $10^{-8}$, we see that in the case of molecular currents always $T_K$ completely dominates -- and therefore all magnetic field lines can penetrate.

Even if we make a ring of a single electron, we cannot imagine to get near the limit of $T_L/T_K=1$, because for that we would require $\log a$ to be of the order of $10^4$.

\vspace{10pt} 

\noindent {\bf 4.} We now want to investigate whether we can reach this limit with a model, that is closer to reality.

Consider a sphere, with radius $a$, over whose volume $N$ electrons are evenly distributed, while the entire sphere rotates around its axis with angular velocity $\Theta$.

For this case, one can find in Abraham~\cite{abraham1903}
the following values for the two contributions to the energy that we want to consider:
\begin{eqnarray}
    T_L & = & \alpha \frac{N^2 e^2 a}{c^2} \Theta^2, \\
    T_K & = & \beta N m a^2 \Theta^2 = 
    \beta \frac{N e^2}{6 \pi R c^2} a^2 \Theta^2,
\end{eqnarray}
in which
\begin{equation}
    \alpha = \frac{2}{5.7}; \; \;
    \beta = \frac{1}{5}.
\end{equation}
The ratio that we are looking for now becomes
\begin{equation}
    \frac{T_L}{T_K} = \frac{6 \pi \alpha}{\beta}   N \frac{R}{a}
    = 5 N \frac{R}{a}.
\end{equation}
Since $N \cong \frac{4}{3} \pi a^3 (10^8)^3$, this becomes
\begin{equation}
    \frac{T_L}{T_K} \cong 2 \cdot 10^{12} a^2.
\end{equation}
If this ratio is near 1, than $a$ must have a value of the order of $10^{-6}$. 

In addition to the case where the charge is distributed throughout the volume, we consider the case where the charge is distributed over the surface of the sphere. For this case, Abraham provides the same formulas for $T_L$ and $T_K$, which only different values for $\alpha$ and $\beta$, namely:
\begin{equation}
    \alpha = \frac{1}{9}; \; \;
    \beta = \frac{1}{3},
\end{equation}
so that now the ratio becomes
\begin{equation}
    \frac{T_L}{T_K}
    = 6 N \frac{R}{a}.
\end{equation}
Again, this makes no significant difference to the previous case. We may trust that a different distribution of the charge throughout the sphere will give a result of the same order of magnitude.

The roughness of our estimates is large, and the model does not correspond to reality. (For example, one must imagine that the electrons have the same linear velocity and not the same angular velocity.) 
Nevertheless, the above analysis suggests that it will be possible to experimentally realize, with superconducting particles, the situation where the influences of $T_L$ and $T_K$ would be comparable.\\

{\em Leiden, November 1925}

\section{Historical context and analysis}
\label{sec:context}

In 1925, the field of physics was still firmly within the realm of ``old" quantum mechanics. The Schrödinger equation and Heisenbergs matrix methods, which started the revolution of ``new'' quantum mechanics, would both be only postulated that year.
It is therefore not surprising that De~Haas-Lorentz models the superconductor in a ``semiclassical" way, i.e., acknowledging somehow the quantum nature of the phenomenon (electrons), yet comparing and calculating energies in a classical fashion. 

This work was likely one of the first theoretical papers attempting to understand the microscopic nature of superconductors.
Back then, very little was known in the field of superconductivity, and very few experimental groups were able to perform such low temperatures experiments. One of those groups was indeed Wander de~Haas' laboratory, who took over the lab of Kamerlingh Onnes himself.

\subsection{Superconductivity before 1925}
The phenomenon of superconductivity was discovered in 1911 by Heike Kamerlingh Onnes in Leiden, The Netherlands, at that time the leading place for low-temperature research. 
The discovery was reported that same year in the first Solvay Conference.
In the following few years before World War I, Kamerlingh Onnes carefully studied the drop of resistivity of mercury; discovered other superconductors such as lead and tin (in 1912), and thallium and indium (in 1919). He also studied the effect of impurities, observing that clean and dirty samples didn't show substantial changes in their superconducting properties\footnote{Thanks to Anderson we now know that this is caused by the fact that {\em non-magnetic} impurities in {\em conventional} ($s$-wave) superconductors cannot break time reversal symmetry, hence preserving superconductiving properties~\cite{anderson1959}.}. 
Most importantly, in 1914 they observed that the supercurrent in a closed superconducting wire was sustained without the need of an electromotive force. 
In 1913 Kamerlingh Onnes started working on Perrin's dream of a superpower magnet, capable of sustaining very strong magnetic fields, exploiting this new superconducting feature.
He discovered some other interesting phenomena associated with superconductivity: 
the observation of critical currents were reported in September 1913, followed by the observation in 1914 of the detrimental effect caused by a transverse magnetic field.
Kamerlingh Onnes also studied the behavior in temperature of this critical field, writing the phenomenological relation 
\begin{equation}
    H_c(T)=H_{c,0}\left[ 1-\left(\frac{T}{T_c}\right)^2 \right].
\end{equation}

The political situation of World War I, when helium was rationed for military purposes, severely limited further superconductivity research the remaining of the 1910's.

When Wander de Haas took over the Kamerlingh Onnes lab in 1924, he continued with the studies of the magnetic properties of superconductors, working on the difference between transverse and longitudinal fields. 

It would take until 1933 before the next big discovery in the field of superconductivity took place: the Meissner effect \cite{meissner1933neuer}.
The first phenomenological understanding of superconductivity is commonly attributed to the London brothers, Fritz and Heinz, in 1935 \cite{London1935}. 

In 1925, there were thus more questions than answers. 

\subsection{Analysis}

This paper is sometimes cited as the first paper in which the London penetration depth is discussed~\cite{Bremmer1936,Hoddeson1987,Fossheim2004}.
This is not entirely correct as there is no explicit discussion of a penetration depth. 
Surface currents are only mentioned as an example of how the perfect diamagnetism of a superconductor can be realized.
However in \cite{Bremmer1936}, Bremmer and de Haas mention that {\em ``the discussions by Mrs. De Haas~\cite{Berta1925-SC} and by Becker, Heller, and Sauter~\cite{becker1933stromverteilung} have shown that these fields can penetrate in bodies, the lengths of which are below about 10$^{-6}$ cm''}\footnote{The explicit estimation of 10$^{-6}$ cm belongs to Ref.~\cite{becker1933stromverteilung}.}. We can assume that this was how the paper was known at that time.
We will elaborate more on the relation to the London penetration depth in Section\,\ref{sec:GL}.

What makes this paper noteworthy, however, is the fact that it was possibly the first theoretical attempt towards a microscopic theory of superconductivity and, more significantly, the first to address the {\bf perfect diamagnetism} of superconductors. 

Unknown to De~Haas-Lorentz at the time, there is a subtle but important difference between a perfect diamagnet ($d\Phi/dt = 0$) and the flux expulsion of the Meissner effect ($\Phi = 0$).

In \cite{London1948}, Fritz London wrote:
\emph{``In fact an equation of this type} [$d\Phi/dt=0$] {\em has been proposed several times as basis of a macroscopic electrodynamics of superconductivity in the sense of describing infinite conductivity \cite{Berta1925-SC}. But after the so-called Meissner-Ochsenfeld effect~\cite{meissner1933neuer} had been discovered it became clear that the assumption leads to a great number of current distributions which cannot be realized within superconductors and
that one has to introduce a supplementary restriction} [$\Phi=0$] {\em in order to obtain only the currents which actually exist."}.
This restriction mentioned by London is addressed in the De Haas-Lorentz paper, although the underlying physical intuition remains uncertain. When solving $d\Phi/dt=0$ in Eq.~\eqref{eq:perfectdiamagnet}, she ``equivalently'' rewrites it as $\Phi=0$ in the subsequent step, thus implicitly falling into the Meissner effect scenario. This step she takes ``as an example'', though, and there seems no physical argument for it in her paper.

In Paragraph {\bf 2.}, the derivation follows the assumption that electrical currents carry energy in their kinetic energy as well as their field.
Whether or not flux is expelled depends on the ratio between kinetic energy $T_K$ and the magnetic/inductance energy $T_L$. If the latter is the largest, then we have flux expulsion.
It is interesting to notice that, as we know now, superconducting currents are not simply carried by electrons.
The ``missing term'' here is the condensation energy, encoded in the global quantum mechanical phase of the wavefunction, this last concept being unknown at that time. It would require the development of the Ginzburg-Landau (GL) theory of superconductivity in 1950 to further understand this~\cite{Ginzburg1950}.

\subsection{Modern view}
\label{sec:GL}
Ironically, De Haas-Lorentz writes that the relevant energy consists of {\em three} parts, but then continues to only discuss {\em two} contributions: the kinetic and magnetic energy. Within the modern framework of the GL phenomenological theory of superconductivity, however, the {\em three} contributions to the energy can be quantified as follows:
\begin{enumerate}
    \item The {\bf kinetic energy} associated with the existence of a screening supercurrent. Following Ref.~\cite{Tinkham.2004}, the kinetic energy is
    \begin{equation}
        \mathcal{F}_{K} = \int d^3x \, \frac{1}{2m^*} \left| \left(\frac{\hbar}{i} {\bf \nabla} - \frac{e^*}{c} {\bf A} \right) \psi (x) \right|^2
    \end{equation}
    where $\psi(x)=|\psi(x)|e^{i\varphi(x)}$ is the condensate order parameter. If we assume a $|\psi|$ to be constant -- relevant for strong type-I superconductors -- this expression can also be rewritten in terms of the {\em supercurrent}
    \begin{equation}
        {\bf J} (x) = \frac{e^*}{m^*} |\psi|^2 \left( \hbar {\bf \nabla} \varphi(x) - \frac{e^*}{c} {\bf A}(x)
        \right).
    \end{equation}
    Using the definition of the {\em London penetration depth}
    \begin{equation}
        \lambda^2 = \frac{m^*}{\mu_0 |\psi|^2 e^{*2}}
    \end{equation}
    we get the elegant expression
    \begin{equation}
        \mathcal{F}_K = \int d^3x \frac{\mu_0}{2} \lambda^2 {\bf J}^2.
        \label{Eq:KineticEnergyGL}
    \end{equation}
    \item The {\bf magnetic energy} associated with having a magnetic field inside the superconductor is given by
    \begin{equation}
        \mathcal{F}_B = \int d^3x \frac{1}{2 \mu_0} {\bf B}^2.
        \label{Eq:MagneticEnergyGL}
    \end{equation}
    \item The contribution that De Haas-Lorentz did not include is the {\bf condensate energy}, 
    \begin{equation}
        \mathcal{F}_{\rm SC} = \int d^3x 
            \left( - \alpha | \psi|^2 + \frac{1}{2} \beta |\psi|^2 \right).
    \end{equation}
    At the critical field $H_c$, the magnetic energy should balance the condensate energy, so that we can also write
    \begin{equation}
        \mathcal{F}_{\rm SC} = - \int d^3 x \frac{1}{2\mu_0} H_c^2.
    \end{equation}
\end{enumerate}

Let us apply this to the situation of a superconducting sphere of radius $a$. In the case that an external magnetic field $B_{\rm applied}$ is completely expelled, in the limit of $\lambda \ll a$ the supercurrent is given by
\begin{equation}
    {\bf J} \approx - \frac{B_{\rm applied}}{\mu_0} \frac{3}{2} \frac{r + \lambda}{r \lambda} e^{(r - a)/\lambda} \sin \theta \; \hat{ \mathbf{\varphi}}
\end{equation}
such that the kinetic energy is 
\begin{equation}
    \mathcal{F}_K = \frac{1}{2\mu_0} B_{\rm applied}^2 3 \pi a^2 \lambda
\end{equation}
whereas the magnetic energy is
\begin{equation}
    \mathcal{F}_{B} = \frac{1}{2 \mu_0} B_{\rm applied}^2 \frac{4}{3} \pi a^3.
\end{equation}
Indeed, consistent with the `case I' of Eq.~\eqref{eq:caseI-SC} introduced by Dr.~De~Haas-Lorentz, the magnetic energy is much larger than the kinetic energy. In fact, up to a geometry-dependent factor, we find
\begin{equation}
    \frac{\mathcal{F}_B}{\mathcal{F}_K} \approx \frac{a}{\lambda} \gg 1
\end{equation}
in the case of flux expulsion. In other words, the ratio $T_L/T_K$ is a direct measure of the London penetration depth! However, this was not anticipated by De~Haas-Lorentz, and unfortunately the role of the screening currents would remain unknown for another ten years.

The semiclassical energy comparison introduced by De~Haas-Lorentz, however, does work when the condensate energy is included: as long as the applied field is smaller than the critical field, the condensate energy would be dominant over both kinetic and magnetic energies.

\subsection{Intermediate state}

The final sentence of the paper contains another interesting intuition, already anticipated in Paragraph {\bf 3.}, where the author proposes the possibility of realizing {\em ``an intermediate situation where the influence of both $T_L$ and $T_K$ can be felt"}. 
Following her reasoning, the possibility to {\em ``experimentally realize, with
superconducting particles, the situation where the influences of
$T_L$ and $T_K$ would be comparable''} means to have an intermediate situation between the two cases discussed in Paragraph {\bf 2.}, respectively Eq. \eqref{eq:caseI-SC} and Eq. \eqref{eq:caseII-m}.
In other words, she suggests the possibility of having a partial penetration of field lines inside a superconductor.

She does not comment further, thus this penetration of field lines could also happen homogeneously, not necessarily as an Abrikosov lattice \cite{abrikosov1957}. 
Nevertheless, the phenomenon whereby field lines penetrate a superconductor were experimentally discovered only in 1935 by Rjabinin and Schubnikow \cite{rjabinin1935magnetic} and it is now known as the \emph{mixed state} of type-II superconductors. Note that there exists also an intermediate state of type-I superconductors, depending on the geometry of the sample. For example, a spherical type-I superconductor will have partial penetration of the magnetic field if $H > \frac{2}{3} H_c$.\cite{Tinkham.2004}

\section{Conclusion}

In conclusion, despite the whole modelization being substantially incorrect in light of our current knowledge, Berta de Haas-Lorentz's paper was the first attempt to model superconductivity in a comprehensive microscopic theory and it contains several interesting intuitions.

It would take several years before having theories on superconductivity, a comprehension of which required “new” quantum mechanics by Heisenberg, Schrodinger and Dirac.
The first {\em phenomenological} theory encoding both types of superconductors will be developed by Ginzburg and Landau in 1950 \cite{Ginzburg1950}, whereas 
the {\em microscopic} theory will be published few years after \cite{Bardeen1955, Cooper1956, BCS1957micro, BCS1957theory}.

\section*{Acknowledgements}

G.V. acknowledges Andrey Varlamov for sharing his vast knowledge on the history of superconductivity.
 This work was funded by the Swiss National Science Foundation (SNSF) via Starting Grant TMSGI2 211296.
\bibliographystyle{elsarticle-num-2} 
\bibliography{refs}



\end{document}